\documentclass[aps, prl, twocolumn, showpacs]{revtex4-1}

\usepackage{graphicx}
\usepackage{epstopdf}

\begin{document} 

\title{Observation of quantum jumps in a superconducting artificial atom}
\author{R. Vijay$^*$}
%\email{rvijay@berkeley.edu}
\author{D. H. Slichter$^*$}
\author{I. Siddiqi}
\affiliation{Quantum Nanoelectronics Laboratory, Department of Physics, University of California, Berkeley CA 94720}

\date{\today}

\begin{abstract}
We continuously measure the state of a superconducting quantum bit coupled to a microwave readout cavity by using a fast, ultralow-noise parametric amplifier. This arrangement allows us to observe quantum jumps between the qubit states in real time, and should enable quantum error correction and feedback\textemdash essential components of quantum information processing.
\end{abstract}


\maketitle

A continuously monitored quantum system prepared in an excited state will decay to its
ground state with an abrupt jump. The jump occurs stochastically on a characteristic time scale $T_{1}$, the lifetime of the excited state, provided the measurement is not too strong. Such quantum jumps, originally envisioned by Bohr, have been observed in trapped atoms and ions \cite{dehmeltprl,winelandprl, toschekprl}, single molecules \cite{brauchlenature}, photons \cite{harochenature1}, single electrons in cyclotrons \cite{gabrielse}, microscopic defects in a Josephson junction \cite{YuPRL20081}, and recently in nuclear \cite{jelezkoscience1} and electron \cite{ataturenature1} spins.  Observation of quantum jumps requires a quantum non-demolition (QND) measurement scheme, that is, one which leaves the system in an eigenstate of the measured observable \cite{braginsky}, thus allowing repeated measurements.  One must also be able to perform the measurements on a timescale much faster than $T_1$ in order to resolve the jumps. All previous experiments have used microscopic quantum degrees of freedom with long relaxation times ($\sim$  ms to s). 

Superconducting quantum bits (qubits) \cite{scqubitsrev2008} exploit macroscopic quantum degrees of freedom in an electrical circuit and typically have much shorter relaxation times ($\sim$ $\mu$s) on account of strong coupling to their environment. However, they are easy to manipulate, tunable and can be mass produced, making them a promising candidate for a scalable quantum computing architecture. Moreover, the drawbacks of short relaxation times can be overcome by using quantum error correction \cite{qec_shor} which requires a fast, high-fidelity measurement scheme. Further, if the measurement is QND, one can use feedback techniques to perform continuous error correction \cite{qec_cont}; to date, though, no suitable measurement scheme has been demonstrated.

In this Letter, we report the first observation of quantum jumps in a superconducting qubit\textemdash a macroscopic quantum system\textemdash by implementing a high-fidelity, QND measurement scheme using a fast, ultralow-noise parametric amplifier \cite{hatridge2010}.  Our experiment uses the circuit quantum electrodynamics (cQED) architecture, where the superconducting qubit is dispersively coupled to a superconducting cavity \cite{cQEDtheory1}, in analogy to an atom in a Fabry-Perot cavity. Probing the qubit-state-dependent cavity frequency implements a continuous, high visibility QND measurement\cite{wallraffvisibility1}. 

Despite successfully demonstrating QND measurement with several kinds of superconducting qubits \cite{cQEDexpt1,HouckPRL20081,fluxonium2009}, cQED implementations with linear cavities have typically suffered from low single-shot fidelity, precluding the observation of quantum jumps. This is primarily due to inefficient amplification of the photons leaving the cavity. The noise added by state-of-the-art cryogenic semiconductor microwave amplifiers is considerably larger than the signal from the cavity, necessitating repeated measurements to resolve the qubit state \cite{wallraffvisibility1}. Using more readout photons can induce qubit state mixing \cite{dresseddephasing}, thus limiting the fidelity. Other high fidelity readout schemes implemented for superconducting qubits are either too slow \cite{transmonCBA1} or are not QND \cite{phasequbitrev}. Josephson parametric amplifiers \cite{castellanos-beltran20081,JPCNature1} with near quantum limited noise performance can potentially enable single shot readout
in the cQED architecture, but most existing designs have an instantaneous 
bandwidth below 1 MHz, too small to enable real time
monitoring of the qubit state. Since superconducting qubit lifetimes are
typically around 1 $\mu$s, one would need a
bandwidth of order 10 MHz to resolve quantum jumps between qubit states with high fidelity. We achieve this by using a low quality factor (Q) nonlinear resonator operated as a parametric amplifier \cite{hatridge2010, epaps1}.  

Our experimental setup, shown schematically in Fig. 1, is anchored to the mixing chamber of a dilution refrigerator at 30 mK.  The superconducting readout cavity (brown) is implemented as a quasi-lumped element linear resonator \cite{kevinosborn} consisting of a meander inductor (L=1.25 nH) in parallel with an interdigitated capacitor (C=575 fF).  A transmon qubit \cite{transmontheory1} (blue, $E_J$=11.4 GHz, $E_C$=280 MHz) is capacitively coupled to the cavity. This arrangement is different than typical cQED setups which use transmission line resonators for the cavity. This design has a smaller footprint and avoids detrimental higher cavity modes \cite{HouckPRL20081}. Probe photons enter from the input port and reflect off the readout cavity, acquiring a phase shift that depends on the qubit state.  These photons then travel through a series of circulators, which allow microwave signals to propagate in one direction as indicated by the arrows in the figure, to the parametric amplifier (paramp), which amplifies the signal and sends it to the output port.  The signal is further amplified by cryogenic and room temperature amplifiers before being mixed down to zero frequency, digitized, and recorded.  

\begin{figure}[tbp]
\begin{center}
\includegraphics[width=89mm]{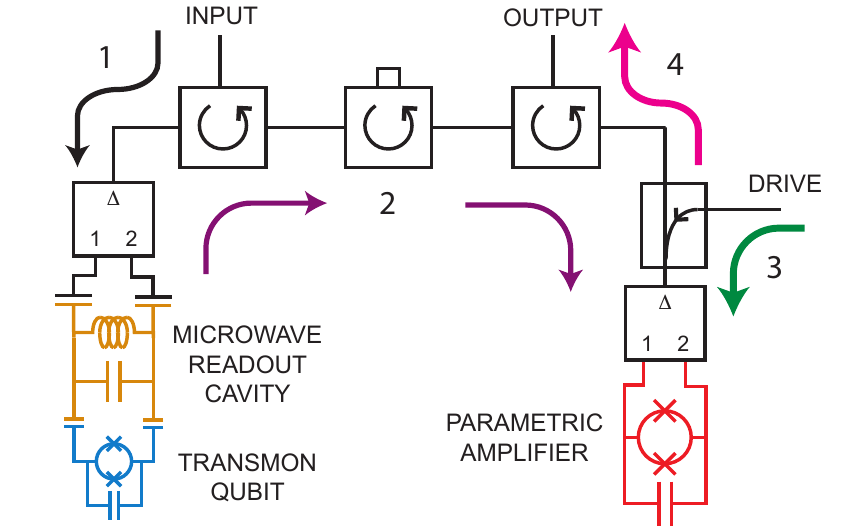}
\end{center}
\caption{Experimental setup (color online).  Readout photons (black arrow, \#1) enter from the input port and are directed through a microwave circulator to a 180$^\circ$ hybrid, which converts the single-ended microwave signal into a differential one.  The photons interact with the readout cavity and the reflected signal (purple arrows, \#2) carries information about the qubit state toward the parametric amplifier (paramp) through three circulators, which isolate the readout and qubit from the strong pump of the paramp.  A directional coupler combines this signal with pump photons (green arrow, \#3) from the drive port.  The paramp amplifies the readout signal, and the amplified signal (magenta arrows, \#4) is reflected and sent through the third circulator to the output port. Qubit manipulation pulses are also sent via the input port.}  
\end{figure}

\begin{figure}[t]
\begin{center}
\includegraphics[width=89mm]{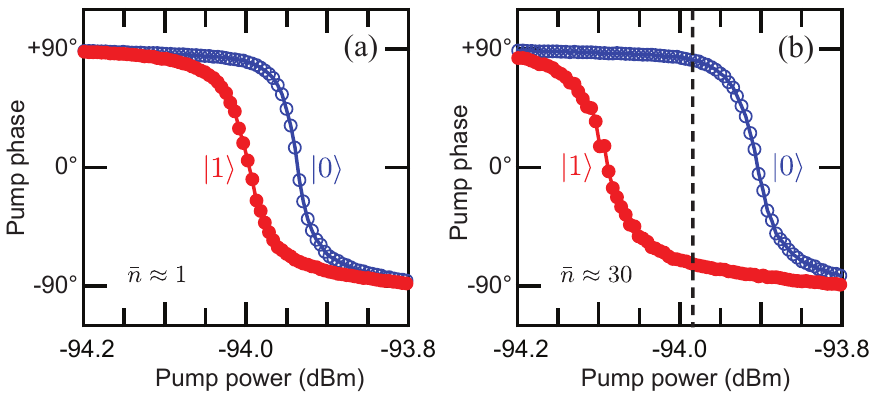}
\end{center}
\caption{Parametric amplifier response (color online).  In (a), we plot the measured average phase of the reflected pump as a function of pump power with the qubit prepared in the ground (blue, open circles) or excited (red, filled circles) states and an average readout cavity occupation of about one photon.  Panel (b) shows the same curves for higher average photon number ($\bar{n}\approx30$) in the readout cavity, which increases the relative separation. The vertical dashed line shows the optimal bias point.}
\end{figure}

\begin{figure*}[t]
\begin{center}
\includegraphics{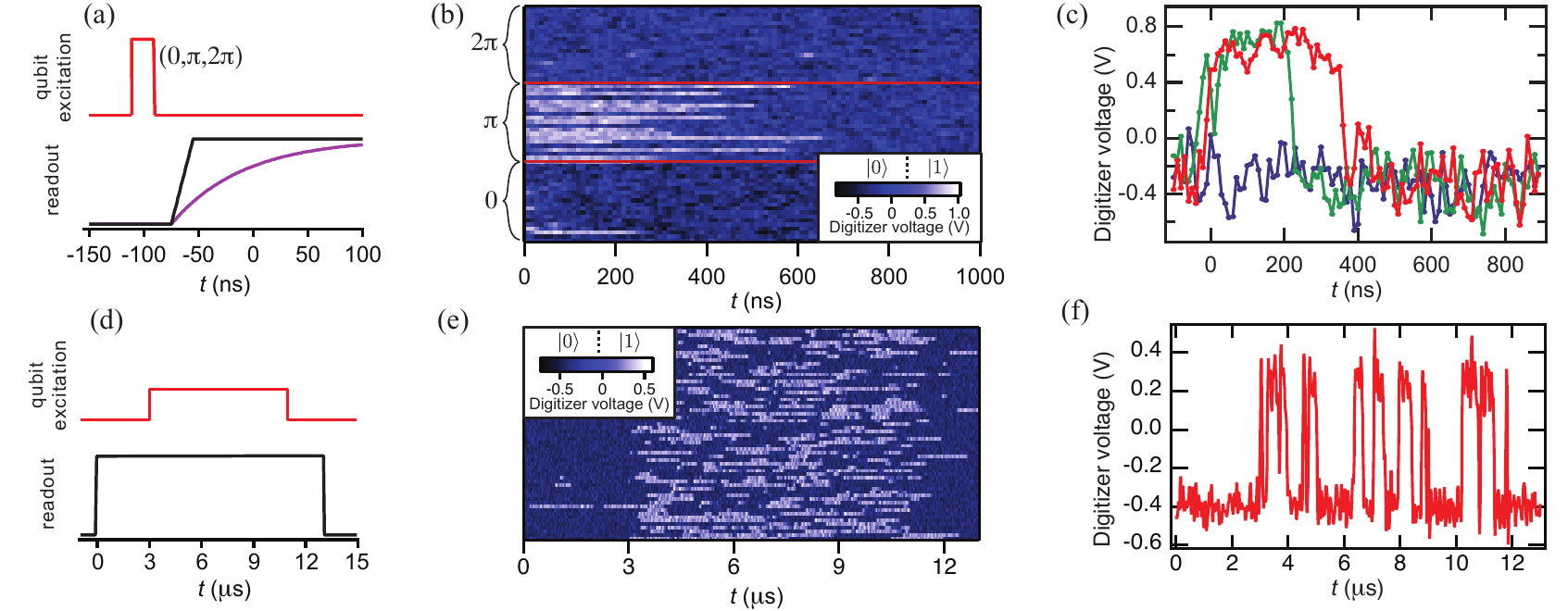}
\end{center}
\caption{Quantum jumps (color online). (a) shows the pulse sequence used to generate (b) and (c).  The qubit is excited with a pulse of varying amplitude (red), and the readout (black) is immediately energized, causing the cavity population (purple) to rise and effect a measurement. Time $t=0$ corresponds to two cavity time constants after the readout is energized. (b) shows 20 representative single-shot traces for three different qubit rotations ($0,\pi$, and $2\pi$).  Abrupt quantum jumps from the excited state (white) to the ground state (blue) are clearly visible for the data corresponding to the $\pi$ pulse, while the traces corresponding to 0 and $2\pi$ indicate that the qubit is mostly in the ground state. (c) shows single-shot time traces following a $\pi$ (red and green) and $2\pi$ (blue) pulse.  (d) shows the pulse sequence used to generate (e) and (f).  Here the readout is energized with the qubit in the ground state, and then a continuous qubit drive is applied after a 3 $\mu$s delay.  (e) and (f) show 60 traces and one trace, respectively, of the qubit jumping between the ground and excited state under the influence of both the qubit drive and measurement pinning.}
\end{figure*}

The bare readout cavity frequency is $5.923$ GHz, with a linewidth $\kappa/2\pi= 4.9$ MHz.  The qubit frequency is set at $4.753$ GHz,
corresponding to a detuning $\Delta/2\pi =1.170$ GHz.  The cavity is driven at $5.932$ GHz (the cavity frequency corresponding to the qubit in the ground state), and we define $\bar{n}$=$(\bar{n}_g+\bar{n}_e)/2$, where $\bar{n}_g$ and $\bar{n}_e$ are the average readout cavity occupations at a given excitation power with the qubit in the ground and excited states, respectively.  These occupations are calibrated using the ac Stark effect for a multilevel qubit \cite{epaps1} .  
The measured coupling strength $g/2\pi=109$ MHz results in a dispersive shift of the cavity \cite{transmontheory1} due to the qubit state $2\chi/2\pi =4.3$ MHz.  We measure qubit lifetime $T_1=320$ ns and dephasing time $T_2^*=290$ ns at this qubit operating point. 

When the paramp is appropriately biased with a strong pump tone, the reflected pump has a power-dependent phase shift as shown in Fig. 2. Since the phase changes sharply with pump power, a small change in the pump power due to an additional signal at the pump frequency is amplified into a large phase shift on the pump signal. This mode of operation only amplifies signals in phase with the pump signal and theoretically adds no noise \cite{Caves1982}, maintaining the signal-to-noise ratio (SNR) of the input signal. If the noise on the input signal is only due to quantum fluctuations, then the noise floor can be expressed as a noise temperature $T_Q=\hbar \omega_s / 2k_B \approx 142 $ mK, where $\omega_s/2 \pi = 5.932$ GHz is the signal frequency.  By contrast, typical microwave amplification chains have a system noise temperature in the range 10 - 30 K, about two orders of magnitude higher.  Near noiseless operation, along with large bandwidth, was previously demonstrated for this paramp design \cite{hatridge2010}.

Since $2\chi\approx\kappa$, measurement photons at $5.932$ GHz exiting the readout cavity will have a relative phase shift of about 180 degrees, depending on the state of the qubit.  When these photons arrive at the paramp, they coherently add to or subtract from the pump (also at $5.932$ GHz, and tuned to be in phase with the readout photons), causing a phase shift of up to 180 degrees in the reflected pump photons which form the output signal.  This can be seen in Fig. 2(a), where the average phase of the output signal is plotted as a function of pump power.  The two traces correspond to measurements taken with the qubit prepared in the ground (blue, open circles) and excited (red, filled circles) states, with $\bar{n}\approx1$ photon.  Increasing the number of photons in the cavity further separates the ground and excited state curves as shown in Fig. 2(b) for $\bar{n}\approx 30$ photons.  This level of excitation of the readout cavity maximizes readout SNR while keeping the measurement QND, and was used for all further measurements discussed below.

With this technique, we can perform single-shot measurements of the qubit state and observe quantum jumps.  We prepare the qubit state with a 20 ns pulse of varying amplitude at the qubit frequency of 4.753 GHz and immediately probe the cavity with photons at 5.932 GHz. The amplified signal is then mixed down to zero frequency, effectively converting the phase shift signal of the readout into a single-quadrature voltage signal.  This voltage is then digitized at 10 ns intervals. Fig. 3(a) shows the pulse sequence and  Fig. 3(b) plots 20 individual traces for each of three pulse amplitudes corresponding to $0,\pi$, and $2\pi$ qubit rotations. We discriminate between ground and excited states using a threshold level of 0.25 V, as indicated on the scale bar. One can clearly see abrupt quantum jumps from the excited state
(white) to the ground state (blue) for the data corresponding to a $\pi$ pulse, while the traces corresponding to $0$ and $2\pi $ show the qubit mostly in the ground state.   A few traces after 0 and $2\pi$ pulses show jumps to the excited state, and a few traces after a $\pi$ pulse are never measured to be in the excited state.  We attribute the first effect to qubit state mixing due to high photon numbers in the readout cavity \cite{dresseddephasing}, and the second effect to the qubit spontaneously decaying before the cavity can ramp up \cite{wallraffvisibility1}.  Three representative traces of the quantum jumps are shown in Fig. 3(c), one where the qubit was prepared in the ground state (blue) and two where it was prepared in the excited state (red and green) and subsequently relaxed to the ground state at different times. The SNR in the measured traces is defined as $\mathrm{SNR}_{\mathrm{meas}}=|\mu_g-\mu_e|/(\sigma_g+\sigma_e)$ where  ($\mu_g,\mu_e$) and ($\sigma_g, \sigma_e$) are the mean and standard deviation of the digitizer voltages corresponding to the ground and excited states respectively. We measure an SNR of about 3.75, a factor of 2.3 lower than the theoretical value of $\sqrt{\bar{n}_{g}\kappa / B}=8.5$, where B=20 MHz is our measurement bandwidth.  We attribute this discrepancy primarily to saturation of the paramp at this high readout power; details are discussed in the supplementary information \cite{epaps1}.

We also investigated the effect of simultaneous qubit excitation and measurement.  We energize the readout and then turn on a long qubit excitation pulse after a few $\mu$s, as shown in Fig. 3(d).  This qubit drive tries to coherently change the qubit state while the projective measurement forces the qubit to be in the ground or excited state, resulting in the random telegraph signal seen in Figs. 3(e) and 3(f).  Note that the discrimination threshold here (-0.05V) is different than that in Figs. 3(b) and (c) due to different bias conditions for the paramp.  Previous measurements \cite{palacios20101} have only been able to indirectly infer such quantum jumps from the averaged spectrum of the measurement signal. This inhibition of qubit state evolution due to measurement is the essence of the quantum Zeno effect \cite{winelandzeno1990,gambettapra20081} and will be the subject of future work using samples with longer coherence times and further improvements in measurement signal-to-noise ratio.  

\begin{figure}[htbp]
\includegraphics[width=89mm]{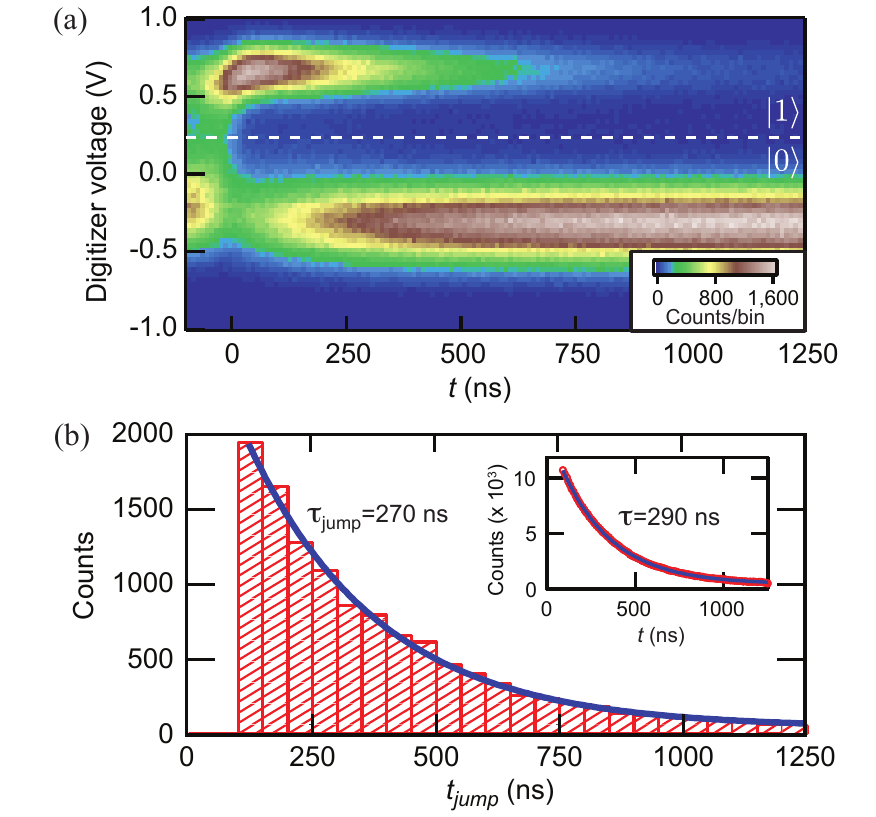}
\caption{Jump statistics (color online). (a) shows a histogram of $2\times 10^4$ individual measurements with the qubit prepared with a $\pi$ pulse and $\bar{n}\approx30$ readout photons. The excited state signal, centered around 0.6 V, is clearly resolved from the ground state signal at -0.3 V; the dashed line shows the discrimination threshold.  The ensemble population is predominately in the excited state at $t=0$, and decays to the ground state with a time constant ((b), inset)  $\tau=290$ ns. (b) shows a histogram of jump times from the excited state to the ground state extracted from individual measurements. The solid line is an exponential fit with a time constant $\tau_{jump}=270$ ns.  We do not plot jumps that occur less than two cavity time constants after the readout is energized.  The pulse protocol used is shown in Fig. 3(a).}
\end{figure}

Finally, we look at the statistics of these quantum jumps. Fig. 4(a) plots a histogram of $2\times 10^4$ individual measurements with the qubit prepared in the excited state, as a function of digitizer voltage and time \cite{epaps2}. Most of the population is measured in the excited state (centered around 0.6 V) at $t=0$ and then decays  to the ground state (centered around -0.3 V) with a time constant $\tau=290$ ns (Fig. 4(b), inset). Despite the large separation between the ground and excited state peaks, the maximum qubit readout fidelity is about $70\%$. This can be almost entirely attributed to the measured $T_1$=320 ns being comparable to the cavity rise time $2/\kappa$=65 ns, which means that around $30\%$ of the excited state population decays to the ground state before the measurement is made. Since we can resolve individual decay events, we can also plot a histogram of the jump (excited to ground state) times \cite{epaps3} as shown in Fig. 4(b). The histogram shows an exponential decay with a time constant $\tau_{jump}=270$ ns. Both these time constants are consistent with each other and with the measured $T_1$ of the qubit, as would be expected for a QND measurement \cite{dresseddephasing}. 

In conclusion, we have demonstrated high-fidelity, real-time monitoring of a superconducting qubit and observed quantum jumps in this macroscopic quantum system. This is the first fruit of a powerful technique for quantum measurements in solid state systems and is a major step toward implementing quantum feedback control and quantum error correction.  Our measurement technique can be readily extended to a variety of other systems of interest, including nitrogen vacancy centers in diamond and lower-dimensional semiconductor systems. Furthermore, our work suggests a route to study the quantum Zeno effect \cite{gambettapra20081} and to shed further light on non-idealities in quantum measurement processes \cite{dresseddephasing}.  The high fidelity quantum measurements we have demonstrated can also be used to realize a time-resolved single microwave photon source/detector \cite{gambettapra20081}, thus enabling a new class of quantum optics experiments in the solid state.

\begin{acknowledgments}
We thank  M. H. Devoret, R. J. Schoelkopf,  J. Gambetta, M. Hatridge and O. Naaman for useful discussions. R.V. and I.S. acknowledge funding from AFOSR under Grant No. FA9550-08-1-0104. D.H.S. acknowledges support from a Hertz Foundation Fellowship endowed by Big George Ventures.
\end{acknowledgments}

\end{document}

\title{Supplement to ``Observation of quantum jumps in a superconducting artificial atom"}
\author{R. Vijay$^*$}

\author{D. H. Slichter$^*$}

\author{I. Siddiqi}
\affiliation{Quantum Nanoelectronics Laboratory, Department of Physics, University of California, Berkeley CA 94720}

\date{\today}

\maketitle

\renewcommand{\thefigure}{S\arabic{figure}}

\section{Photon number calibration}

To calibrate the photon occupation of the readout cavity, we use the ac Stark effect \cite{schusteracstark} for a multilevel qubit \cite{gambettaprl2010}, which accounts for higher order nonlinearities.  We performed qubit spectroscopy in the presence of a measurement drive at $5.932$ GHz, close to the resonant frequency of the cavity with the qubit in the ground state (5.933 GHz).  For each measurement power, we used the shift in the qubit frequency to determine $\bar{n}_g$, the cavity occupation with the qubit in the ground state.  If the qubit is in the excited state, the cavity resonant frequency is detuned from the drive frequency, and thus the cavity occupation $\bar{n}_e$ is estimated to be lower by a factor of 3.5.  For data shown in Figs. 3 and 4, we estimate $\bar{n}_g\approx47$ and $\bar{n}_e\approx13$.  The number $\bar{n}\approx30$ quoted in the main text is the average of these two.  Note that when investigating quantum Zeno or dressed dephasing effects, it is important to account for the different cavity occupations between the excited and ground states.  The measurement frequency can also be chosen to be midway between the excited state and ground state cavity resonant frequencies, so that the photon population in the cavity will be the same for both qubit states at a given drive power. 
 
\begin{figure}[htb]
\begin{center}
\includegraphics[width=89mm]{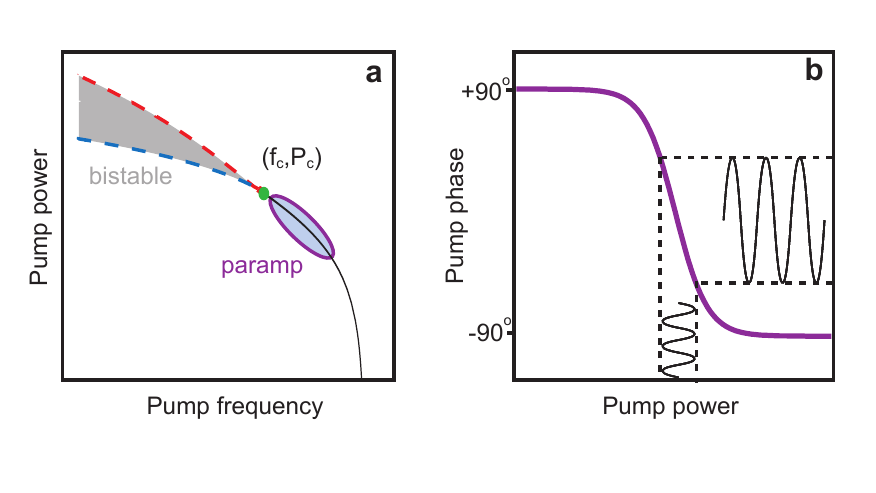}
\caption{The phase diagram of the amplifier resonator is shown schematically in \textbf{(a)}.  The resonance frequency (solid line) gradually decreases with increasing drive power.  Above a critical drive power $P_C$ and below a critical frequency $f_C$ the resonator is bistable.  At powers slightly below $P_C$, the resonator functions as a parametric amplifier (paramp).  \textbf{(b)} shows a linecut of \textbf{(a)} at constant pump frequency in the paramp regime.  A weak signal modulating the drive power leads to large changes in the reflected phase of the pump.}
\end{center}
\end{figure}

\section{Amplification mechanism}

The phase diagram of our non-linear resonator is shown schematically in Fig. S1a. The solid line indicates the effective resonant frequency as a function of bias frequency and power. For drive power $P>P_{C}$ and drive frequency $f<f_{C}$, the system is bistable within the boundary indicated by the dotted lines.  This bistability has been employed for qubit readout in the past\cite{vijayRSI,transmonCBA}.  In this work, we bias the resonator just outside the bistable regime. This bias region, marked as ``paramp" in Fig. S1a, has been accessed to demonstrate parametric amplification\cite{castellanos-beltran2007,Vijay2008a, hatridge2010}. Fig. S1b explains the principle of small signal degenerate parametric amplification.   The reflected pump phase is plotted as a function of pump power, giving the transfer function of the amplifier. A weak input signal at the pump frequency is combined with the pump and modulates its amplitude, resulting in a large change in the phase of the reflected pump.  Since there are many more pump photons than signal photons, this effectively amplifies the input signal.  

Two key points should be noted here. First, this technique implements a
phase-sensitive amplifier \cite{Caves1982}, since only signals in phase with the
pump will be amplified. Signals which are 90 degrees out of phase with the pump do not affect the pump amplitude to first order and hence do not get amplified. Moreover, in this phase-sensitive mode, the paramp amplifies signal and noise by the same factor.  Since the signal to noise ratio (SNR) is not degraded, the amplification process is noiseless \cite{hatridge2010}.  Secondly, the amplifier is linear only for input signals
which keep the bias conditions in the linear region of the transfer function
shown in Fig. S1b.  However, in this work we are only concerned with two possible cavity states
corresponding to the two states of the qubit.  Therefore, the linearity of the amplifier
is inconsequential here as long as it faithfully maps the two cavity states
to two distinct output states. Unfortunately, as will be discussed later, operation in the saturated regime is no longer noiseless.   It is clear from Fig. 2 of the main text that the input signal to the paramp, even for $\bar{n}\approx1$ in the readout cavity, is large enough to saturate the paramp. Note that the paramp can be used as a phase insensitive amplifier as well when the signal is detuned from the pump \cite{hatridge2010}. In this case, the ideal paramp adds the minimum noise required by quantum mechanics\textemdash half a photon at the signal frequency.

\section{Signal to Noise ratio estimation}

Here we discuss the expected signal-to-noise ratio (SNR) for resolving qubit states using an ideal paramp and compare it with the SNR we achieved experimentally.  We need to calculate the intrinsic SNR of two coherent states of the same amplitude but differing in phase by 180 degrees. If the average photon occupation of a cavity driven at resonance is $\bar{n}_\mathrm{res}$, then it can be shown that the power radiating out of the cavity is $P_{\mathrm{rad}}=\bar{n}_\mathrm{res} \hbar \omega_s \kappa$, where $\omega_s$ is the signal frequency and $\kappa$ is the cavity linewidth.  When the cavity is driven in the reflection geometry, the steady state incident power is the same as the steady state reflected power which forms the output. This output power is $P_{\mathrm{out}}=P_{\mathrm{rad}}/4$ where the factor of four arises due to the interference between the incident and radiated signal. The root mean square (rms) voltage corresponding to this power on a 50 ohm transmission line is $V_{\mathrm{out}}=\sqrt{50 P_{\mathrm{out}} }$. The intrinsic noise floor is set by quantum fluctuations and can be expressed as a noise temperature $T_Q=\hbar \omega_s /2 k_B$. The corresponding voltage noise is given by $V_Q=\sqrt{50 k_B T_Q B}$, where B is the measurement bandwidth. If this signal is sent through an amplification chain with a system noise temperature $T_{\mathrm{sys}}$, then the effective voltage noise referred to the input of the chain is given by $V_{\mathrm{sys}}=\sqrt{50 k_B (T_Q+T_{\mathrm{sys}}) B}$. We can then define the SNR for resolving two coherent states of amplitude $\sqrt{2} V_{\mathrm{out}}$ and differing in phase by 180 degrees as SNR $= \sqrt{2} V_{\mathrm{out}} / V_{\mathrm{sys}}$. In the case of noiseless amplification, $T_{\mathrm{sys}}=0$, and  SNR$=\sqrt{\bar{n}_\mathrm{res}\kappa / B}$. Putting in the experimental parameters  $\kappa/2\pi = 4.9$ MHz, $B=20$ MHz, we get an SNR $ \approx 1.24 \sqrt{\bar{n}_\mathrm{res}}$. As expected, the SNR can be improved by increasing $\bar{n}_\mathrm{res}$, but in practice higher order effects start to affect the dispersive shift \cite{gambettaprl2010} and cause qubit state mixing \cite{dresseddephasing}, setting an upper bound on $\bar{n}_\mathrm{res}$ for our measurement.  In our experiment, we used a maximum photon occupation of $\bar{n}_\mathrm{res}=\bar{n}_g=47$ as described above. For this level of excitation, we compute an optimal SNR $\approx 8.5$.

\begin{figure}[htb]
\begin{center}
\includegraphics[width=89mm]{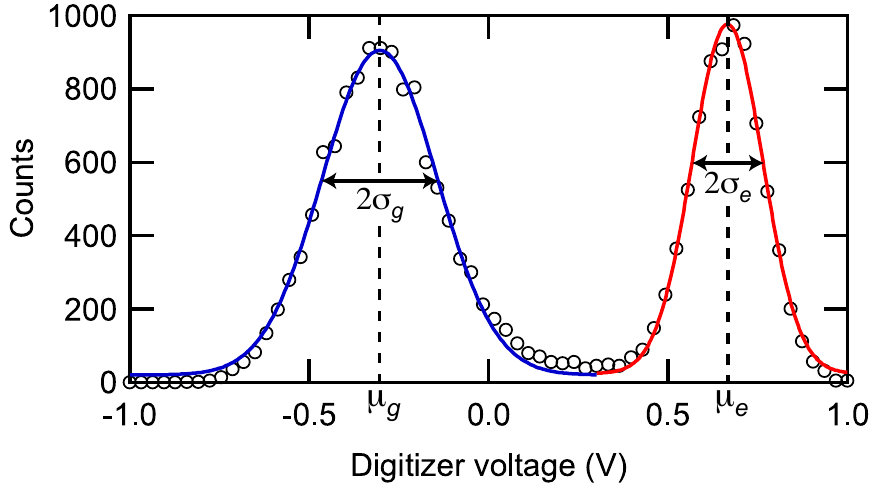}
\caption{Output histograms for computing SNR.  The data (open circles) are a vertical linecut of Fig. 4a at time $t=200$ ns.  The solid lines are Gaussian fits to the peaks corresponding to ground (blue) and excited (red) states.  The means ($\mu_g,\mu_e$) and standard deviations ($\sigma_g, \sigma_e$) of each fit are shown.}
\end{center}
\end{figure}

Fig. S2 shows a voltage histogram of 2$\times 10^4$ individual traces for $\bar{n}_\mathrm{res}=47$; the data are from a vertical linecut of Fig. 4a at $t=200$ ns.  There are two clearly resolved peaks, corresponding to the qubit in ground (negative voltage) and excited (positive voltage) states.  We can fit each peak to a Gaussian function (blue and red lines) and extract the means ($\mu_g,\mu_e$) and standard deviations ($\sigma_g, \sigma_e$) for each fit.   Note that the widths of the two peaks are different because the paramp is not linear for this high level of excitation.  We define the measured SNR as $\mathrm{SNR}_{\mathrm{meas}}=|\mu_g-\mu_e|/(\sigma_g+\sigma_e)$. Using numerical simulations, we have verified that this definition of SNR is consistent with the analytical expression given in the previous paragraph.  For the data shown $\mathrm{SNR}_{\mathrm{meas}} \approx 3.75$, a factor of 2.3 smaller than the ideal case. This implies that $T_\mathrm{sys} \approx 4.1\, T_Q =0.59$ K and the added noise is not zero. Nevertheless, this is an improvement of nearly two orders of magnitude in noise temperature over a typical state-of-the-art microwave amplification chain using cryogenic semiconductor amplifiers, where $T_{\mathrm{sys}}\approx 10 - 30$ K. 

We attribute the discrepancy between the predicted and measured SNR to three sources. We believe the dominant contribution is due to saturation of the paramp, which in our case occurs for cavity occupations above $\bar{n}\approx1$. This is evidenced by the steady deterioration of the system noise temperature with increasing photon occupation in the cavity.  For the data in Fig. S2,  $\bar{n}_\mathrm{res}=47$ and the paramp is deeply in the saturated regime.  In the linear regime, the output SNR grows linearly with the input SNR from the cavity; the output signal grows, while the output noise stays constant.  When the paramp is saturated, the amplitude of the output signal no longer grows with increasing input signal amplitude.  Input noise continues to be amplified, although the noise gain decreases gradually as the paramp moves farther into the saturated regime.  As a result, the output SNR plateaus and grows only sub-linearly with increasing input SNR; the output signal is fixed, while the output noise is slowly decreasing.  Thus when the paramp is deeply saturated it degrades the SNR and is no longer noiseless. Nevertheless, the added noise of the paramp is still much lower than that of a typical amplification chain.  It is important to note as well that the output SNR is still a monotonically increasing function of input SNR, so that operation in the saturated regime is advantageous. Improving the dynamic range of the paramp will prevent saturation and should allow further improvements in SNR. 

A second contribution to the SNR discrepancy is insertion loss between the cavity and the paramp, primarily due to losses in the three circulators use to isolate the cavity from the strong pump of the paramp. This could be reduced by integrating the paramp on chip. The third source of discrepancy is the noise temperature of the amplification chain following the paramp, which in our case was about 25 K.  This noise is comparable in magnitude to the quantum noise of the readout signal after amplification by the paramp, and so raises the system noise temperature.  We can address this effect by lowering the noise temperature of the following amplifier chain, for example by using a lower noise semiconductor amplifier or by reducing the cable losses between the paramp and the semiconductor amplifier.  

\section{Sample fabrication and characterization}

The qubit and readout resonator were fabricated on a bare high-resistivity Si wafer using electron beam lithography and double-angle Al evaporation with an intervening oxidation step.  The evaporated films were 35 and 80 nm thick respectively and were deposited with an e-beam evaporator.  The amplifier resonator was fabricated on a high-resistivity Si wafer.  The wafer was patterned with a rectangular 300-nm-thick sputtered Nb ground plane and coated with 125 nm of SiN$_x$ grown by PECVD.  The amplifier resonator was then fabricated on top of the SiN$_x$ layer using electron beam lithography and double-angle Al evaporation with oxidation, in the same manner as the readout resonator \cite{hatridge2010}. 

The experiment was carried out in a dilution refrigerator with a base temperature of 30 mK.  The readout resonator had inductance L=1.25 nH and capacitance C=575 fF, with a bare resonant frequency of 5.923 GHz and a linewidth of 4.9 MHz.  These numbers and the behavior of the resonator match well with finite element simulations performed using Microwave Office.  The amplifier resonator was formed by a two junction superconducting quantum interference device (SQUID) with a maximum critical current of 4 $\mu$A shunted by two parallel plate capacitors in series with a total capacitance of 6.5 pF, giving a resonant frequency tunable between 4 and 6.2 GHz and a loaded Q (quality factor) of about 25.  The transmon qubit had measured parameters $E_J=11.4$ GHz and $E_C=280$ MHz at the operating frequency of 4.753 GHz, corresponding to a maximum $E_J$ of 17.9 GHz at zero flux bias.  The qubit transition frequency could be tuned between 4 and 5.8 GHz.  Measurements of $T_1$ as a function of qubit frequency agreed well with decay due to the single-mode Purcell effect in combination with an unknown relaxation channel giving an effective Q of 11,500 \cite{HouckPRL2008}.  Ramsey fringes yielded $T_2= 290$ ns at the operating point.  We observed a small leakage of pump photons from the amplifier resonator back to the readout resonator, consistent with the finite isolation of the circulators.  Using the ac Stark effect, we determined the leakage corresponded to a readout cavity population of much less than one photon.  The leakage did cause dephasing, which we verified by Ramsey fringes.  We were able to cancel this leakage by applying a coherent microwave tone to the input port.  Because the amplifier resonator has a very fast response, one could avoid the leakage problem by turning the paramp off during qubit manipulation and evolution.